\documentclass[aps,prmaterials,longbibliography,superscriptaddress,amssymb,reprint,10pt]{revtex4-2}

\usepackage{graphicx}
\usepackage{dcolumn}
\usepackage{bm}
\usepackage{color}
\usepackage{amsmath,amsfonts,amssymb}
\usepackage{graphicx}
\usepackage{siunitx}
\usepackage{booktabs,longtable}
\usepackage{natmove}
\usepackage[version=4]{mhchem}
\usepackage{hyperref}

\newcommand{\avg}[1]{\left<{#1}\right>}

\let\oldcite\cite
\renewcommand*\cite[1]{ \oldcite{#1}}
\DeclareMathOperator{\real}{Re}
\DeclareMathOperator{\imag}{Im}

\begin{document}

\author{Jamal Ben Youssef}
\email{jamal.ben-youssef@univ-brest.fr}
\author{Nathan Beaulieu}
\affiliation{LabSTICC, CNRS, Universit{\'{e}} de Bretagne Occidentale, 29238 Brest, France}
\author{Richard Schlitz}
\affiliation{Department of Physics, University of Konstanz, 78457 Konstanz, Germany}
\author{Davit Petrosyan}
\affiliation{Department of Materials, ETH Zurich, H{\"{o}}nggerbergring 64, 8093 Zurich, Switzerland}
\author{Michaela Lammel}
\affiliation{Department of Physics, University of Konstanz, 78457 Konstanz, Germany}
\author{William Legrand}
\email{william.legrand@neel.cnrs.fr}
\affiliation{Institut N{\'{e}}el, CNRS, Universit{\'{e}} Grenoble Alpes, 38042 Grenoble, France}

\title{Low-temperature-compatible iron garnet films grown by liquid phase epitaxy}

\begin{abstract}
Single-crystalline yttrium iron garnet (YIG) thin films ($<$~\SI{100}{\nano\meter}) form the backbone of magnonics, owing to the record-low losses affecting their magnetization dynamics. However, thin epitaxial YIG has mostly been investigated under ambient temperatures, limited by the paramagnetic losses occurring at low temperatures due to the gadolinium gallium garnet (GGG) substrates required for epitaxial growth. Driven by a growing interest in magnonic devices that can operate in cryogenic conditions and address quantum information applications, there is a strong need for iron garnet epitaxial films grown on diamagnetic substrates that can maintain low losses at low temperatures. In this work, we use liquid phase epitaxy (LPE) to grow ultrathin films of strained YIG on a commercial diamagnetic substrate, yttrium scandium gallium garnet (YSGG). We investigate their magnetization dynamics in the 3--\SI{300}{\kelvin} temperature range, and compare them to equivalent films grown on paramagnetic GGG. We demonstrate for LPE YIG on YSGG substrates a ferromagnetic resonance linewidth below \SI{1}{\milli\tesla} at \SI{3}{\kelvin}, together with a very weak temperature and frequency dependence of the losses. The growth of YIG/YSGG by LPE provides a straightforward approach to producing iron garnet thin films for use in low-temperature investigations.
\end{abstract}

\maketitle

\section{Introduction}

In recent years, the idea of exploiting the quantum properties of exchanged-coupled large scale spin ensembles has been conceptualized as an evolution of the field of magnonics\cite{Tabuchi2016,Yuan2022}. It constitutes a very actively pursued direction in theoretical works, e.g.,\cite{Li2018a,Kamra2019,Yuan2020,Sharma2021,Sun2021a,Falch2025arXiv}, that still has to be paired with a matching level of experimental exploration. This motivates the search for magnetically ordered solid-state systems with low magnetic dissipation, able to host magnons with measurable quantum coherence. So far, advanced experiments in quantum magnonics have relied on bulk, single-crystalline spheres of yttrium iron garnet (YIG), a magnetic insulator, as the magnon host material\cite{Tabuchi2015,LachanceQuirion2020,Wolski2020,Xu2023a,Rani2025}. Developing alternatives to bulk spheres\cite{Spencer1959,Serha2025arXiv} in the form of films that would have equivalent magnon coherence would not only introduce the possibility of varying the structure of the magnon modes at play, but could also enable a scaling toward magnonic micro- or even nano-devices, to be integrated with other quantum systems\cite{Fukami2021,Li2022a}. 

Achieving the longest magnonic lifetimes, corresponding primarily to narrowest resonance linewidths, requires the highest degree of spatial homogeneity for the magnonic system, best achieved in single-crystalline systems. Therefore, YIG thin films addressing the requirements of high-end magnonics demand an epitaxial growth on garnet substrates with a matching lattice parameter, which is traditionally chosen to be gadolinium gallium garnet (GGG). Unfortunately, this precludes their use for integrated magnonics operating under cryogenic conditions, because the high rare-earth content of GGG makes it strongly paramagnetic. The paramagnetic \ce{Gd^{3+}} cations from the GGG substrate couple by interfacial exchange mechanisms and dipolar interactions to the magnetic \ce{Fe^{3+}} cations in YIG, which causes a rapid degradation of the magnonic lifetimes with decreasing temperature\cite{Jermain2017}, down to the \si{\milli\kelvin} range\cite{Kosen2019,Knauer2023,Serha2024,Schmoll2025}. It has been demonstrated that films grown by liquid phase epitaxy (LPE) are able to mitigate this degradation of the linewidth for the homogeneous ferromagnetic resonance (FMR) mode, owing to a minimal interdiffusion of cations at the interface between the film and the substrate\cite{Beaulieu2018,Will-Cole2023}. Nevertheless, the dipolar coupling mechanisms affecting magnons with non-zero wavevectors remain active\cite{Knauer2023,Schmoll2025}, as well as other inductive couplings with any integrated microwave circuitry. At the present stage, substrate paramagnetism represents a longstanding issue that needs to be solved to enable further progress toward low-temperature magnonics. Whereas several techniques enabling a separation of the YIG films from their GGG growth substrate\cite{Rachford2000,Baity2021,Xu2025} are appealing in spite of their added complexity, replacing GGG with a non-paramagnetic substrate appears as an ideal approach in most situations.

Less than a dozen of compositions of garnet substrates that can be batch-produced by the Czochralski method are commonly commercialized. Among them, only yttrium scandium gallium garnet (YSGG)\cite{Brandle1973} is diamagnetic and features a lattice parameter close enough to that of YIG so it can suitably replace GGG. Consequently, recent efforts have been expanded on replacing GGG by YSGG substrates for the epitaxy of YIG thin films devoted to low temperatures\cite{Guo2022,Guo2023,Legrand2025a}. Low-temperature FMR measurements have confirmed the suitability of the YIG/YSGG pairing to reduce magnon losses in cryogenic conditions. In these previous studies, the growth relied on magnetron sputtering deposition, which is particularly prone to deviations from ideal film stoichiometry and to an excessive cationic interdiffusion at the substrate/film interface. Such aspects are necessarily limiting the linewidths that can be achieved both under ambient and cryogenic conditions, compared to the intrinsic linewidth of bulk single-crystalline YIG\cite{Spencer1959,Maier-Flaig2017}. In order to circumvent these issues, we address in this work the growth of YIG thin films on YSGG by LPE.

We first investigate experimentally the regime of growth for strained YIG on YSGG by LPE, and observe the onset of strain relaxation for films thicker than about \SI{100}{\nano\meter}. Structural characterization by X-ray diffraction confirms the fully strained growth of about \SI{60}{\nano\meter}-thick YIG films on YSGG. We then investigate the magnetization dynamics losses in these films by broadband FMR, comparing samples grown on GGG and YSGG. Room-temperature measurements confirm the excellent magnetic quality of these films on both substrates. We then perform temperature-dependent broadband FMR measurements to extract the magnetization dynamics parameters, in particular linewidth, within the 3--\SI{300}{\kelvin} range. Compared to GGG, the YIG films grown on YSGG exhibit a very weak temperature dependence for the FMR linewidth, under both in-plane and out-of-plane applied magnetic fields. The results on two YIG/YSGG films with different strain and room-temperature linewidth tend to link strain and roughness to additional losses in cryogenic conditions. The best results for the present unpatterned ultrathin film samples correspond to a linewidth $\mu_0\Delta{}H=$~\SI{0.7}{\milli\tesla} at \SI{46}{\giga\hertz}, and a Gilbert damping $\alpha<2\cdot10^{-4}$. Ultrathin YIG films grown on commercially available YSGG substrates eliminate the issues associated with substrate paramagnetism in GGG. This approach enables the investigation of low-loss coherent magnonics at cryogenic temperatures, therefore providing exciting prospects for future works in quantum magnonics.

\section{Growth of YIG/GGG and YIG/YSGG by LPE}
\label{sec:model}

The epitaxial thin films of YIG investigated here were grown by LPE from a flux of \ce{PbO} and \ce{B_2O_3}, using high-purity (\SI{99.999}{\percent} or better) sources for the oxides constitutive of YIG. The growth has been performed on either \qtyproduct{10x10}{\milli\meter} or 1-inch-diameter, (111)-oriented substrates of GGG and YSGG. In order to achieve slow growth rates, relevant for thin films in the range of tens of \si{\nano\meter} to around \SI{100}{\nano\meter}, a growth temperature very close to the saturation temperature of the solution has been chosen, corresponding to a supercooling small enough to achieve growth rates of around \SI{1}{\nano\meter\per\second}. Other important aspects concern the control of the rotational speed of the substrate during the growth process and the precise determination of the optimal composition for the melt\cite{Blank1972,Dubs2017,Beaulieu2018,Dubs2020}.

\begin{figure}[t]
\includegraphics[width=3.5in, trim= 0cm 0cm 0cm 0cm]{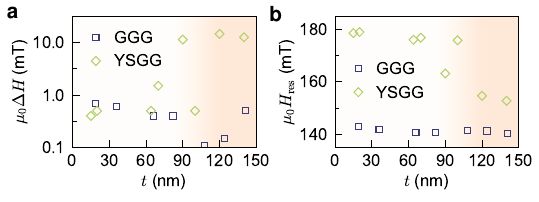}
\caption{In-plane ferromagnetic resonance (IP FMR) characterization at room temperature for a thickness series. At \SI{6}{\giga\hertz}, (a) FMR linewidth $\mu_0\Delta{}H$ in log scale and (b) resonance field $\mu_0H_{\rm{res}}$ against YIG thickness $t$, for films grown on GGG and YSGG. The gradient fill symbolizes the transition to a growth with relaxed strain for YIG/YSGG.}
\label{fig:FMR_strain_RT}
\end{figure}

The growth of YIG on YSGG needs to accommodate a significant lattice parameter mismatch (\SI{-0.7}{\percent}), and is thus prone to strain relaxation by the formation of dislocation defects. This has limited the range of thicknesses that can be grown with other techniques as well, including notably pulsed laser deposition and magnetron sputtering\cite{Gueckelhorn2021,Guo2023}. Proceeding with a series of samples with increasing thickness, all grown from an identical melt, the onset of strain relaxation in YIG grown on YSGG by LPE beyond a limit thickness is evident. For this series of samples, we have measured the room-temperature FMR at \SI{6}{\giga\hertz}, which we can compare to samples grown on GGG as a proxy for sample quality. The FMR linewidth is shown in Fig.~\ref{fig:FMR_strain_RT}a, which evidences a transition toward large linewidths for YIG/YSGG with YIG thicknesses above roughly 90--\SI{110}{\nano\meter}. Accordingly, the resonance field at \SI{6}{\giga\hertz} (Fig.~\ref{fig:FMR_strain_RT}b) reduces when the strain begins to relax, due to the reduction of the magneto-elastic contribution to the magnetic anisotropy field. In the following, we focus on perfectly epitaxial YIG films grown on YSGG for thicknesses below the occurrence of strain relaxation, and compare them to nearly unstrained films grown on GGG. 

\begin{figure*}[t]
\includegraphics[width=7.0in, trim= 0cm 0cm 0cm 0cm]{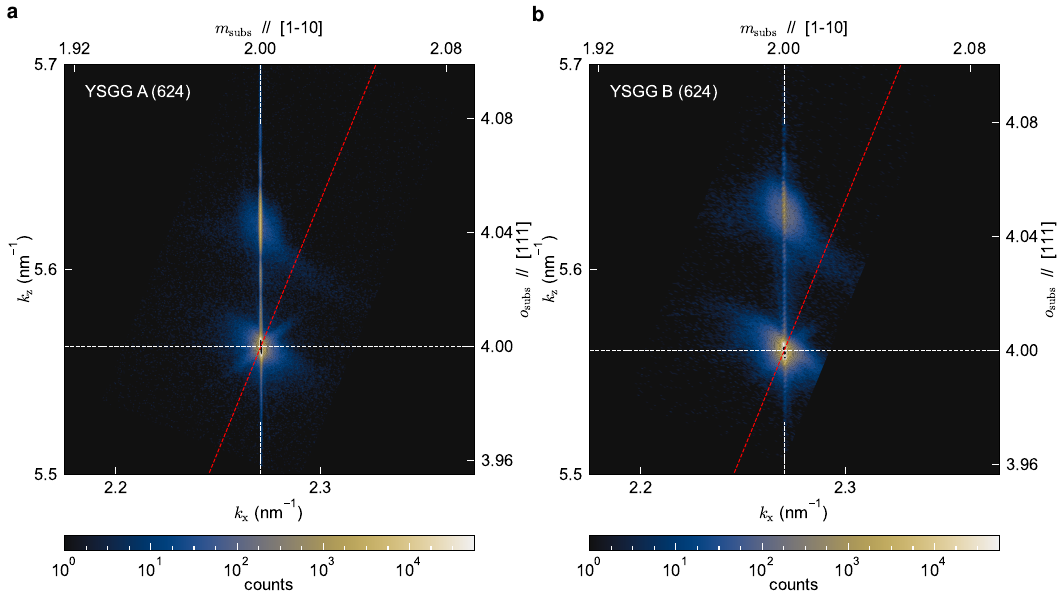}
\caption{X-ray reciprocal space maps around the (624) substrate peaks for (a) YSGG A and (b) YSGG B samples. Vertical and horizontal white dashed lines indicate alignment with substrate (624) peak within the film plane and along the film normal. Red dashed lines indicate the peak position expected in case of relaxed growth. In each panel, the film peaks are aligned with the substrate peak at a same $k_x$, indicating a fully strained growth.}
\label{fig:RSM}
\end{figure*}

The four samples further investigated in this work have been characterized by X-ray reflectivity and X-ray diffraction (data shown in Appendix~\ref{app:XRRD}), with the main sample parameters listed in Table~\ref{tab:samples}. To date, the LPE growth of strained YIG on YSGG is very unusual in comparison to the standard growth of YIG on GGG, and little is known on the structural properties of such films. To confirm the fully strained growth of YIG/YSGG even for thicknesses as large as \SI{60}{\nano\meter}, we have performed reciprocal space map analysis of the lattice relationships within the two samples grown on YSGG. The maps recorded in the vicinity of the (624) peak are shown in Fig.~\ref{fig:RSM}, which exhibit in each case a perfect alignment between the in-plane lattice vectors of substrate and film. Therefore, samples YSGG A and B are free of structural relaxation, and the whole YIG layer maintains a pseudomorphic rhombohedral distortion to match with the lattice parameter of the substrate.

\begin{table*}[t] %or S[table-format=3.1(1),table-space-text-pre=$-$,table-number-alignment=center]
    \caption{Samples investigated in this work, indicating for each: the substrate size used for the growth (circular of 1-inch diameter or square of $10\times10$~\si{\milli\meter\squared}), the YIG thickness $t$, the XRR surface roughness $\sigma$, and substrate and strained film out-of-plane lattice parameters; for in-plane FMR and then for out-of-plane FMR at room temperature, the gyromagnetic ratio $\gamma$, the effective anisotropy field $\mu_0H_{\rm{eff}}$, the apparent Gilbert damping $\alpha'$ and inhomogeneous linewidth $\mu_0\Delta{}H_0$.}
    \footnotesize
    \begin{tabular*}{\textwidth}{l@{\extracolsep{\fill}}ccS[table-format=3.0]S[table-format=1.1]S[table-format=1.4]S[table-format=1.4]S[table-format=2.1]S[table-format=3.0]S[table-format=1.1]S[table-format=1.1]S[table-format=2.1]S[table-format=3.0]S[table-format=1.1]S[table-format=1.1]}
	\hline\hline
	\vphantom{\large{Y}}
	& & & \multicolumn{4}{c}{{Structure}} & \multicolumn{4}{c}{{IP FMR, \SI{293}{\kelvin}}} & \multicolumn{4}{c}{{OOP FMR, \SI{293}{\kelvin}}} \\
	\cline{4-7}\cline{8-11}\cline{12-15}
	\vphantom{\Large{Y}}
	& & & {$t$} & {$\sigma$} & {$a_{\rm{subs}}$} & {$a_{\perp}$} & {$\gamma/(2\pi)$} & {$\mu_0H_{\rm{eff}}^{\rm{ip}}$} & {$\alpha'$} & {$\mu_0\Delta{}H_0$} & {$\gamma/(2\pi)$} & {$\mu_0H_{\rm{eff}}^{\rm{oop}}$} & {$\alpha'$} & {$\mu_0\Delta{}H_0$} \\
	\vphantom{\large{Y}}{System} & {Sample} & {Growth} & {(\si{\nano\meter})} & {(\si{\nano\meter})} & {(\si{\nano\meter})} & {(\si{\nano\meter})} & {(\si{\giga\hertz\per\tesla})} & {(\si{\milli\tesla})} & {($10^{-4}$)} & {(\si{\milli\tesla})} & {(\si{\giga\hertz\per\tesla})} & {(\si{\milli\tesla})} & {($10^{-4}$)} & {(\si{\milli\tesla})} \\
        \hline
	\vphantom{\large{Y}}YIG/GGG & A & 1-in. & 116 & 0.6 & 1.2382 & {-} & 27.9 & 182 & 3.9 & 0.2 & 27.9 & 171 & 1.9 & 0.5 \\
	\vphantom{\large{Y}}YIG/GGG & B & 10\,$\times$\,10 & 82 & {-} & 1.2382 & {-} & 27.7 & 187 & 2.2 & 0.1 & 27.7 & 176 & 2.9 & 0.2 \\
	\vphantom{\large{Y}}YIG/YSGG & A & 10\,$\times$\,10 & 59 & 0.8 & 1.2455 & 1.2321 & 27.7 & 84 & 1.0 & 0.5 & 27.6 & 76 & 1.8 & 0.9 \\
	\vphantom{\large{Y}}YIG/YSGG & B & 1-in. & 64 & 3.4 & 1.2460 & 1.2309 & 27.9 & 70 & 1.5 & 0.6 & 27.9 & 62& 1.8 & 0.3\\
    \end{tabular*}
    \label{tab:samples}
\end{table*}

\begin{figure}[t]
\includegraphics[width=3.5in, trim= 0cm 0cm 0cm 0cm]{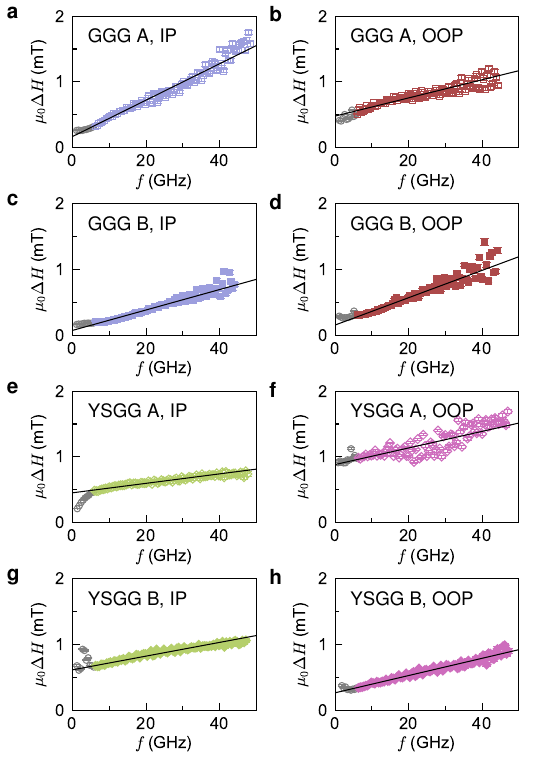}
\caption{Ferromagnetic resonance (FMR) (a,c,e,g) in-plane (IP) and (b,d,f,h) out-of-plane (OOP) linewidth $\mu_0\Delta{}H$ against frequency $f$, at \SI{293}{K}, for films (a,b) YIG/GGG A, (c,d) YIG/GGG B, (e,f) YIG/YSGG A and (g,h) YIG/YSGG B. Solid lines are fits to $\mu_0\Delta{}H=\mu_0\Delta{}H_0+4\pi\alpha'{}f/\gamma$, gray symbols show the low-frequency data points excluded from these fits.}
\label{fig:FMR_RT}
\end{figure}

The magnetization dynamics parameters of these samples have been verified by FMR, first at room temperature. The grown samples are diced in smaller rectangular pieces (typically 2--\SI{4}{\milli\meter} along each direction), placed on a broadband coplanar waveguide excited at varied frequencies, and aligned in the field of an electromagnet in order to measure the dynamical magnetic susceptibility of the films against the external static magnetic field, applied either in the film plane (IP) or perpendicular to the plane (OOP). The resonance fields $\mu_0H_{\rm{res}}(f)$ are fit with the Kittel law $2\pi{}f=\gamma\mu_0[(H_{\rm{res}}+H_{\rm{eff}}^{\rm{ip}})H_{\rm{res}}]^{1/2}$ (IP), $2\pi{}f=\gamma\mu_0[H_{\rm{res}}-H_{\rm{eff}}^{\rm{oop}}]$ (OOP), with $\gamma$ the gyromagnetic ratio and $\mu_0H_{\rm{eff}}^{\rm{ip,oop}}$ the effective anisotropy field acting in each case, including the magnetization component $\mu_0M_{\rm{s}}$ as well as uniaxial and magneto-crystalline anisotropy terms. The value of $\mu_0H_{\rm{eff}}$ varies slightly between IP and OOP, because the magneto-crystalline contribution to the effective anisotropy is different for these different crystalline orientations. The resonance linewidths $\mu_0\Delta{}H(f)$ are fit with the Gilbert model $\mu_0\Delta{}H=4\pi\alpha'{}f/\gamma+\mu_0\Delta{}H_0$, where $\alpha'$ is an apparent value for the dimensionless Gilbert damping parameter and $\mu_0\Delta{}H_0$ is the apparent inhomogeneous linewidth. This analysis is simplified with respect to the variety of possible magnetization dynamics dissipation mechanisms, which do not all have a linear dependence on frequency. However, the $\mu_0\Delta{}H(f)$ curves appear close to linear in the films investigated here, at least beyond \SI{6}{\giga\hertz}. Because the damping mechanisms outside of the Gilbert framework only have a weak influence on the present films, retaining only $\alpha'$ and $\mu_0\Delta{}H_0$ as effective parameters to describe the linewidth behavior constitute a practical simplification. However, we highlight that $\alpha'$ does not exactly correspond to the intrinsic value of Gilbert damping $\alpha$ that would be found in the very high frequency limit. The key magnetic parameters are summarized in Table~\ref{tab:samples}, and the data for $\mu_0\Delta{}H(f)$ both IP and OOP are shown in Fig.~\ref{fig:FMR_RT}. All four samples display very good magnetization dynamics at room temperature. However, owing to differences in their substrate geometry and material, they show different dissipation behaviors, which proves insightful for the measurements that follow.

\vspace{-6pt}
\section{Temperature-dependent FMR characterization of YIG/GGG versus YIG/YSGG}
\label{sec:cryoFMR}

We have characterized the temperature-dependent dynamical magnetic properties of these films of YIG/YSGG and YIG/GGG, in the temperature range 3--\SI{300}{\kelvin}, by broadband FMR using a vector network analyzer (VNA) setup. The samples are aligned within the main field directions of a two-axis superconducting magnet in order to perform FMR with external static magnetic field applied either IP or OOP. Within the film plane, the external static magnetic field is applied transverse to the coplanar waveguide probing line, in order to minimize stray field inhomogeneities due to the paramagnetism of the GGG substrates\cite{Serha2024,Serha2025}. At some defined frequencies, standing waves appear due to reflections from microwave components in the setup, which affects phase and amplitude of the microwave signal \cite{Legrand2025} and may modulate the waveguide radiation losses. Other procedure details and representative curves are provided in Appendix~\ref{app:microwaves}. 

\begin{figure*}[t]
\includegraphics[width=7.0in, trim= 0cm 0cm 0cm 0cm]{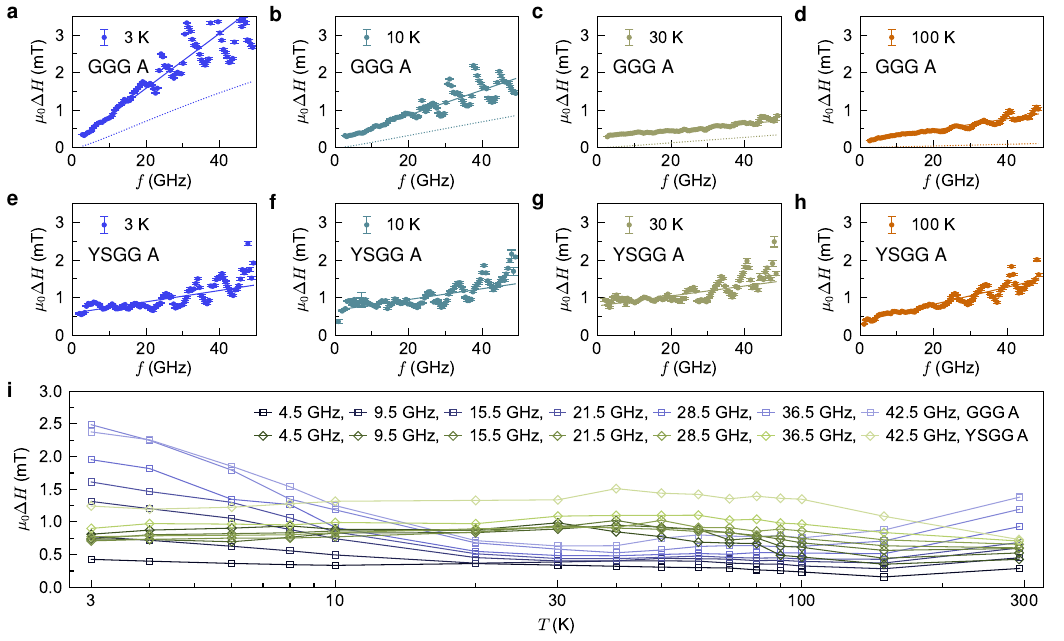}
\caption{In-plane ferromagnetic resonance (IP FMR) losses in the temperature range 3--\SI{300}{\kelvin} for films YIG/GGG A and YIG/YSGG A. Resonance linewidth $\mu_0\Delta{}H$ against frequency $f$ for films grown on (a--d) GGG and (e--h) YSGG, at (a,e) \SI{3}{K}, (b,f) \SI{10}{K}, (c,g) \SI{30}{K}, and (d,h) \SI{100}{K}. Solid lines are fits to $\mu_0\Delta{}H=\mu_0\Delta{}H_0+4\pi\alpha'{}f/\gamma$, and in panels a--d the dotted lines show the expected contribution to the FMR linewidth of the substrate stray field inhomogeneity $\Delta{}B_{\rm{stray}}$. (i) FMR linewidth $\mu_0\Delta{}H$ at several frequencies for films YIG/GGG A and YIG/YSGG A, against temperature.}
\label{fig:comparegood}
\end{figure*}

We start by comparing the properties of the single-crystalline YIG thin films grown on a GGG substrate versus YSGG substrate, samples YIG/GGG A and YIG/YSGG A with a YIG thickness of \SI{116}{\nano\meter} and \SI{59}{\nano\meter}, respectively. Their respective microwave transmission data at a few representative frequencies are provided in Appendix~\ref{app:microwaves}. The frequency-dependent linewidth at several temperatures for these two samples are shown in Fig.~\ref{fig:comparegood}a--h, and the temperature dependence of their FMR linewidth at a few selected frequencies is summarized in Fig.~\ref{fig:comparegood}i. The most striking difference between the two substrates is that the linewidth of YIG/YSGG shows almost no evolution within the temperature range 3--\SI{100}{\kelvin}, while for YIG/GGG the linewidth deteriorates considerably in cryogenic conditions due to the paramagnetism of GGG. This showcases the improvement obtained by using a diamagnetic substrate\cite{Guo2023,Legrand2025a}, instead of commonly used GGG.

\begin{figure}
\includegraphics[width=3.5in, trim= 0cm 0cm 0cm 0cm]{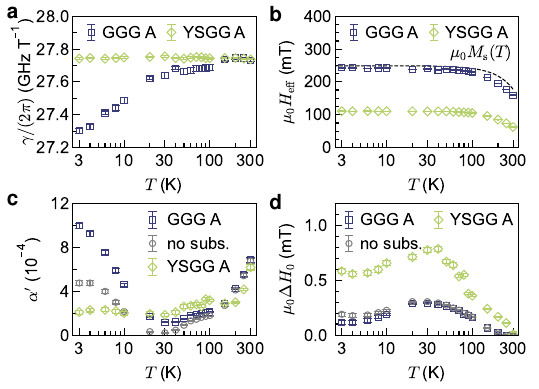}
\caption{Comparison of magnetization dynamics parameters obtained for IP FMR on films YIG/GGG A versus YIG/YSGG A, against temperature $T$. (a) Gyromagnetic ratio $\gamma/(2\pi)$, (b) effective anisotropy $\mu_0H_{\rm{eff}}$, (c) apparent Gilbert damping $\alpha'$, (d) inhomogeneous broadening $\mu_0\Delta{}H_0$. $M_{\rm{s}}(T)$ of YIG is reported as a dashed line in panel b. In panels c and d, the gray symbols labeled `no subs.' show $\alpha'$ and $\mu_0\Delta{}H_0$ deduced after removing the substrate stray field contribution $\Delta{}B_{\rm{stray}}(f)$ from $\mu_0\Delta{}H(f)$ of YIG/GGG A.}
\label{fig:comparegood_summary}
\end{figure}

The temperature-dependent magnetization dynamics parameters for these two samples are summarized in Fig.~\ref{fig:comparegood_summary}. The evolution of the gyromagnetic ratio $\gamma$ with temperature (Fig.~\ref{fig:comparegood_summary}a) highlights the influence of the GGG substrate paramagnetism. Due to the in-plane external field $\mu_0H$ setting the resonance, the GGG substrate acquires a magnetization that is mainly collinear with the external field. This magnetization of the substrate generates a stray field $\mu_0H_{\rm{stray}}$ within the YIG film, mainly along a direction opposite to the external field. This additional field causes an apparent offset of the external field required to reach resonance, manifesting as a change of effective $\gamma$. This effect is absent for YIG films on YSGG. The evolution of $\mu_0H_{\rm{eff}}$ with temperature (Fig.~\ref{fig:comparegood_summary}b) mainly reflects the evolution of the saturation magnetization of YIG, which is indicated by the black dashed line in the plot. A difference of \SI{132}{\milli\tesla} is found between the effective anisotropy fields at low temperatures for the two films of YIG grown on GGG and on YSGG. This corresponds to the magneto-elastic contribution originating in the tensile strain of YIG grown on YSGG. The slight deviation of $\mu_0H_{\rm{eff}}(T)$ from the trend of $\mu_0M_{\rm{s}}(T)$ is due to the evolution of the other contributions to the effective anisotropy in the magnetic system, including both the magneto-elasticity and the cubic anisotropy of YIG. We now move on to the discussion of the temperature-dependent FMR linewidth data on the two substrates, first on GGG and then on YSGG.

Even when the external magnetic field is applied transverse to the probing line in order to minimize edge effects, the stray field resulting from the substrate paramagnetism of GGG remains significantly inhomogeneous\cite{Serha2024}, and causes a broadening of the resonance by an extra linewidth contribution\cite{Serha2025} that we designate $\Delta{}B_{\rm{stray}}$. At a given temperature, this inhomogeneous stray field is proportional to the magnetization in the GGG substrate. Therefore, it increases with frequency as the field required to reach resonance, following the inverse function of the Kittel law. It is also inversely proportional to the temperature, until gradually reaching the magnetic saturation of the GGG at low temperatures and large fields\cite{Serha2024}. When approaching saturation, the linear magnetic susceptibility $\chi_{\rm{p}}$ of GGG becomes insufficient to model its magnetic response. The substrate magnetization and the $\Delta{}B_{\rm{stray}}$ contribution, across all field and temperature regimes that are relevant to the present analysis, are modeled in several simple steps provided in Appendix~\ref{app:GGGfield}. The graph provided in Fig.~\ref{fig:GGGprop}f of Appendix~\ref{app:GGGfield} reveals that this extrinsic contribution accounts for most of the linewidth appearing in Fig.~\ref{fig:comparegood}a,b, with an estimated $\Delta{}B_{\rm{stray}}\approx$ \SI{1.8}{\milli\tesla} at \SI{50}{\giga\hertz} and \SI{3}{\kelvin}, as well as \SI{0.9}{\milli\tesla} at \SI{50}{\giga\hertz} and \SI{10}{\kelvin}. Once this extrinsic contribution is subtracted from $\mu_0\Delta{}H(f)$ at each temperature, we obtain values for $\alpha'$ and $\mu_0\Delta{}H_0$ that are corrected for substrate stray field effects, shown as gray symbols in Fig.~\ref{fig:comparegood_summary}c,d. While $\mu_0\Delta{}H_0(T)$ is nearly unaffected after the subtraction, $\alpha'(T)$ passes through a minimum around \SI{30}{\kelvin}, and features a much more moderate increase with decreasing temperature in the 3--\SI{20}{\kelvin} range. We deduce from the present analysis that the increase in FMR linewidth for YIG/GGG under cryogenic conditions, visible in Fig.~\ref{fig:comparegood}i, is essentially due to the increase in apparent $\alpha'$ coming from the stray fields of the GGG substrate. Other mechanisms related to the gadolinium content at the film/substrate interface, such as interfacial coupling to the \ce{Gd^{3+}} moments\cite{Roos2022} or slow-relaxer impurity behaviors\cite{Dillon1959,Dillon1962,Seiden1964}, are expected to be active at low temperatures. However, they represent less detrimental contributions to the FMR linewidth of thin LPE YIG\cite{Will-Cole2023}, as we confirm here. Those other contributions induced by gadolinium can also be avoided, provided that substrate paramagnetism is eliminated.

For YIG grown on YSGG, some qualitatively different contributions to the FMR linewidth are observed. The data summarized in Fig.~\ref{fig:comparegood_summary}c,d reveals that with decreasing temperature, the apparent damping contribution stabilizes below \SI{30}{\kelvin} at $\alpha'\approx2\cdot10^{-4}$, not far from its level for YIG/GGG when substrate paramagnetism is excluded. By contrast, the inhomogeneous linewidth term peaks near $\mu_0\Delta{}H_0\approx$~\SI{0.8}{\milli\tesla} at 30--\SI{40}{\kelvin}, before reducing to $\approx$~\SI{0.6}{\milli\tesla} at \SI{3}{\kelvin}, which is significantly higher than on GGG. This more prominent inhomogeneous broadening can be assigned to the strain state of YIG grown on YSGG. To a lesser extent, it could also originate in the lower thickness of the specific sample used here (imposed by the limit thickness to avoid strain relaxation), as lower thicknesses tend to reinforce the impact of inhomogenities.

\begin{figure*}[t]
\includegraphics[width=7.0in, trim= 0cm 0cm 0cm 0cm]{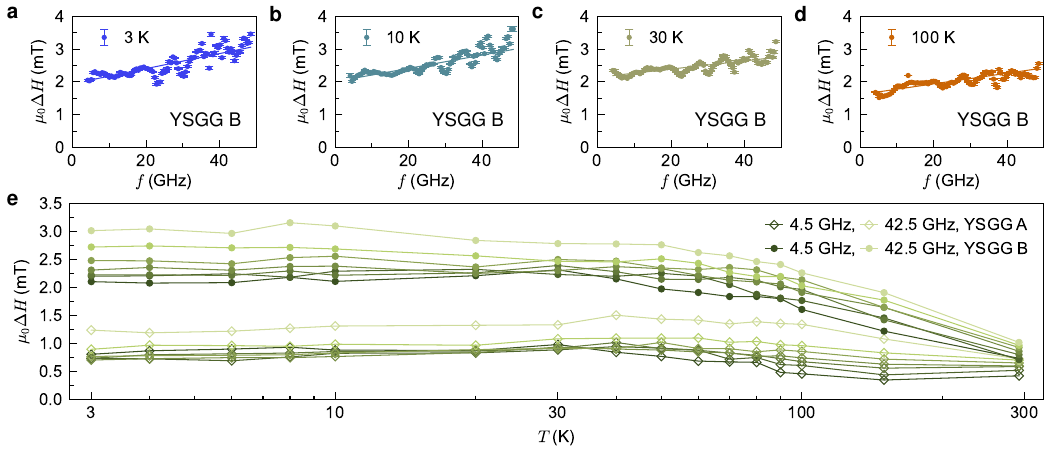}
\caption{In-plane ferromagnetic resonance (IP FMR) losses in the temperature range 3--\SI{300}{\kelvin} for films of YIG/YSGG A and B. (a--d) Resonance linewidth $\mu_0\Delta{}H$ against frequency $f$ for film YIG/YSGG B, at (a) \SI{3}{K}, (b) \SI{10}{K}, (c) \SI{30}{K}, and (d) \SI{100}{K} (data for YIG/YSGG A presented in Fig.~\ref{fig:comparegood}e--h). Lines are fits to $\mu_0\Delta{}H=\mu_0\Delta{}H_0+4\pi\alpha'{}f/\gamma$. (e) FMR linewidth $\mu_0\Delta{}H$ at several frequencies for films YIG/YSGG A and B, against temperature.}
\label{fig:compareYSGG}
\end{figure*}

\begin{figure}[t]
\includegraphics[width=3.5in, trim= 0cm 0cm 0cm 0cm]{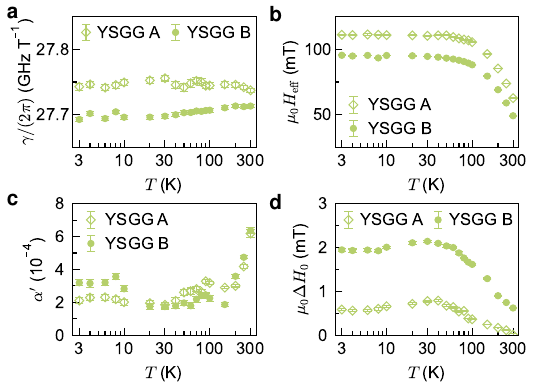}
\caption{Comparison of magnetization dynamics parameters obtained for IP FMR on films of YIG/YSGG A and B, against temperature $T$. (a) Gyromagnetic ratio $\gamma/(2\pi)$, (b) effective anisotropy $\mu_0H_{\rm{eff}}$, (c) apparent Gilbert damping $\alpha'$, (d) inhomogeneous broadening $\mu_0\Delta{}H_0$.}
\label{fig:compareYSGG_summary}
\end{figure}

\section{Investigation of relaxation mechanisms for YIG/YSGG}
\label{sec:complementFMR}

We now aim to investigate more deeply the processes that contribute to the FMR linewidth of YIG/YSGG under cryogenic conditions. For this, we first analyze the FMR results acquired for another LPE sample grown on YSGG, identified as YIG/YSGG B. Its temperature-dependent $\mu_0\Delta{}H(f)$ data is shown in Fig.~\ref{fig:compareYSGG}, and its magnetization dynamics parameters are summarized in Fig.~\ref{fig:compareYSGG_summary}. YIG/YSGG B is grown from the same melt, and has a similar thickness (\SI{64}{\nano\meter}) compared to YIG/YSGG A (\SI{59}{\nano\meter}), but is grown on a slightly different substrate, sourced from a different manufacturer. Due to a difference in the substrate lattice parameter, YIG/YSGG B is slightly more strained than YIG/YSGG A, which is confirmed by the further offset towards small values of the $M_{\rm{eff}}(T)$ curve compared to $M_{\rm{s}}(T)$ (Fig.~\ref{fig:compareYSGG_summary}b). The Gilbert damping part of the linewidth is barely affected by the change of substrate, except for small differences below \SI{10}{\kelvin} (Fig.~\ref{fig:compareYSGG_summary}c). By contrast, the inhomogeneous broadening for YIG/YSGG B is much more pronounced than for YIG/YSGG A (Fig.~\ref{fig:compareYSGG_summary}d). We assign this increase to the larger strain level, which might reach close the maximal level of strain that can be retained for the present \SI{60}{\nano\meter}-thick growth. Hence, achieving best cryogenic FMR linewidths will not only require the use of substrates that are free of paramagnetic cations, but will also require strategies to mitigate the strain and the rhombohedral distortion of the originally cubic unit cell of YIG, which appear to be detrimental to the linewidth.

\begin{figure*}[t]
\includegraphics[width=7.0in, trim= 0cm 0cm 0cm 0cm]{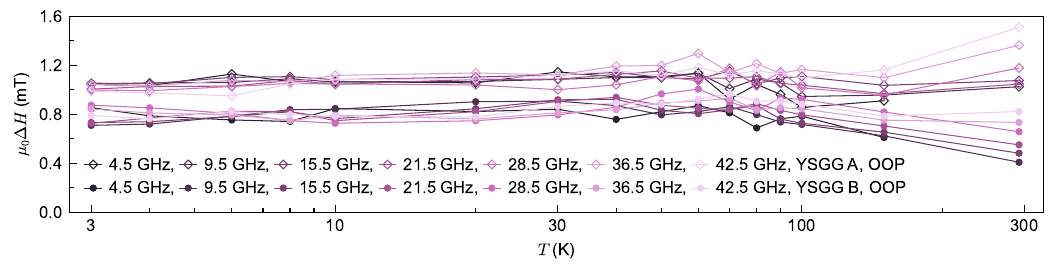}
\caption{Out-of-plane ferromagnetic resonance (OOP FMR) losses in the temperature range 3--\SI{300}{\kelvin} for films of YIG/YSGG A and B, with FMR linewidth $\mu_0\Delta{}H$ at several frequencies against temperature.}
\label{fig:compareYSGGOOP}
\end{figure*}

So far our investigation has focused on IP FMR. This configuration is sometimes preferred for epitaxial garnet films, as it tends to minimize the impact of sample inhomogeneity on $\mu_0\Delta{}H_0$. However, this IP configuration is also prone to an enhanced dissipation of the magnetization dynamics, because of the activation of two-magnon scattering processes\cite{Arias1999a,Lenz2006,Jermain2016}. In addition, the IP FMR of (111)-oriented samples is convenient to avoid the cubic magnetic anisotropy of YIG, which has constant energy in the sample plane defined by the [1-10] and [11-2] directions, but those do not necessarily correspond to the directions with lowest dissipation. It has been shown that for unstrained bulk samples at \SI{4}{\kelvin}, the maximal dissipation is found for magnetic field applied along the [110] directions, while the minimal dissipation is found along the [100] directions\cite{Dillon1958}. The contributions of various impurities to the linewidth vary depending on the measurement direction\cite{Spencer1964,Hartwick1969}. Hence, we expect a very different linewidth behavior for the OOP FMR with external field aligned along the [111] growth direction, compared to the IP FMR in the present films.

\begin{figure}[t]
\includegraphics[width=3.5in, trim= 0cm 0cm 0cm 0cm]{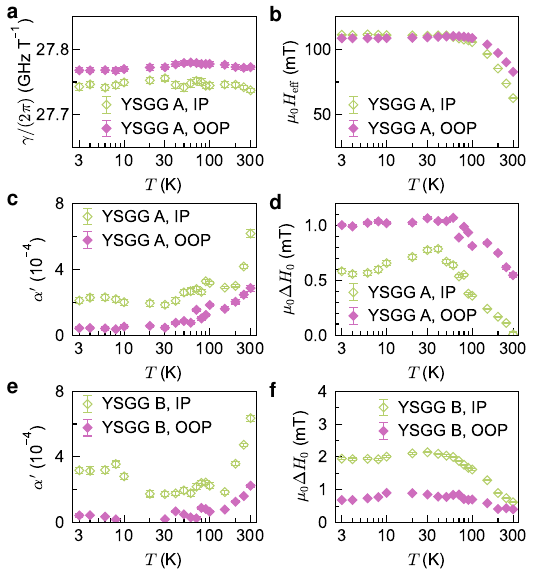}
\caption{Comparison of magnetization dynamics parameters obtained for IP FMR and OOP FMR, on films of YIG/YSGG A and B, against temperature $T$. For YIG/YSGG A, (a) gyromagnetic ratio $\gamma/(2\pi)$, (b) effective anisotropy $\mu_0H_{\rm{eff}}$, (c) apparent Gilbert damping $\alpha'$, (d) inhomogeneous broadening $\mu_0\Delta{}H_0$; for YIG/YSGG B, (e) apparent Gilbert damping $\alpha'$, (f) inhomogeneous broadening $\mu_0\Delta{}H_0$.}
\label{fig:compareYSGGOOP_summary}
\end{figure}

\begin{figure}[t]
\includegraphics[width=3.5in, trim= 0cm 0cm 0cm 0cm]{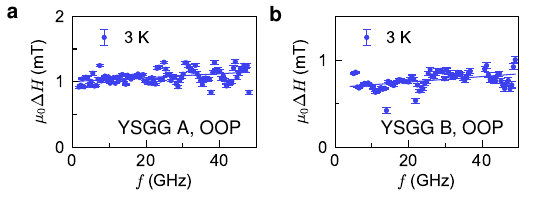}
\caption{Out-of-plane ferromagnetic resonance (OOP FMR) linewidth $\mu_0\Delta{}H$ against frequency $f$, at \SI{3}{K}, for films (a) YIG/YSGG A, and (b) YIG/YSGG B. Solid lines are fits to $\mu_0\Delta{}H=\mu_0\Delta{}H_0+4\pi\alpha'{}f/\gamma$.}
\label{fig:bestYSGGOOP}
\end{figure}

The temperature dependence of the OOP FMR linewidth at a few selected frequencies for YIG/YSGG A and B is summarized in Fig.~\ref{fig:compareYSGGOOP}, and their magnetization dynamics parameters are summarized in Fig.~\ref{fig:compareYSGGOOP_summary}. We highlight their OOP $\mu_0\Delta{}H(f)$ at \SI{3}{\kelvin} in Fig.~\ref{fig:bestYSGGOOP}. In the OOP configuration, the apparent Gilbert damping component of the linewidth (Fig.~\ref{fig:compareYSGGOOP_summary}c,e) is found to drop below $1\cdot10^{-4}$ at low temperatures. At temperatures $T>$~\SI{100}{\kelvin}, it is also significantly reduced compared to the IP configuration. However, a significant inhomogeneous broadening (Fig.~\ref{fig:compareYSGGOOP_summary}d,f) is observed in both YIG/YSGG A and B. When $\mu_0\Delta{}H_0$ dominates in the $\mu_0\Delta{}H(f)$ data as is the case here, the actual damping contribution is very often masked by the large inhomogeneous broadening. This is because their respective contributions to the linewidth are associated to a Lorentzian (Gilbert damping) and likely a Gaussian (inhomogeneous broadening) lineshape, so that they do not simply add up when they have a very different value\cite{McMichael2008}. Still, we can use the data at the highest frequencies to place an upper bound on the intrinsic Gilbert damping $\alpha$ value. For the data in Fig.~\ref{fig:bestYSGGOOP}b, a linewidth $\mu_0\Delta{}H=$~\SI{0.7}{\milli\tesla} is measured at \SI{46}{\giga\hertz}, giving a conservative estimate $\alpha<2\cdot10^{-4}$. We note that the films measured in our study are unpatterned, and thus incorporate a significant inhomogeneous broadening contribution that is inherent to extended LPE films, compared to lithographed micropatterns. Those values are sufficiently low to enable solid future perspectives for cryogenic integrated magnonic devices.

The temperature dependence of $\mu_0\Delta{}H_0$ for both YIG/GGG and YIG/YSGG (Figs.~\ref{fig:comparegood_summary}d and \ref{fig:compareYSGG_summary}d) present a common feature, with a peak around 30--\SI{40}{\kelvin}, which then reduces by $\approx$~\SI{0.2}{\milli\tesla} at the lowest temperatures. Accordingly, the $\mu_0\Delta{}H(T)$ data hints at a relatively small peak in linewidth against $T$, shifting to higher values of $T$ at higher frequencies. We attribute this observation to impurities in the YIG films. Rare-earth impurities are unavoidably present in the source materials, and are thus expected to contribute to the linewidth by a slow-relaxer mechanism\cite{Dillon1959,Dillon1962,Seiden1964}. According to the estimation of Ref.~\oldcite{Spencer1959}, the corresponding total content of rare-earth impurities in the present films shall be of the order of a few \si{ppm} to explain this contribution. Alternatively, a deviation from ideally trivalent iron cations would cause a similar linewidth maximum around \SI{40}{\kelvin}\cite{Spencer1961}. According to the levels measured in Ref.~\oldcite{Hartwick1969}, the total amount of non-\ce{Fe^{3+}} iron cations corresponding to the present peak height would be about 0.02 per formula unit, which corresponds to the typical levels of \ce{Pb^{2+}} inclusion from the flux used to grow YIG by LPE. Although a slight relaxation seems to arise from such impurities, it is smaller for OOP FMR (Figs.~\ref{fig:compareYSGGOOP} and \ref{fig:compareYSGGOOP_summary}d,f) than for IP FMR (Figs.~\ref{fig:compareYSGG}e and \ref{fig:compareYSGG_summary}d). We note that at \SI{3}{\kelvin} this contribution appears almost negligible compared to the total linewidth, which rules out slow-relaxer rare-earth impurities as the main contribution to the cryogenic linewidth observed in the present YIG/YSGG LPE films.

An important observation is that YIG/YSGG A features a much better linewidth than YIG/YSGG B in the IP configuration (Fig.~\ref{fig:compareYSGG}e), whereas it is the opposite in the OOP configuration (Fig.~\ref{fig:compareYSGGOOP}). We ascribe these results to a complex interplay between impurity mechanisms, two-magnon scattering, inhomogeneous broadening and intrinsic damping along different directions in the strained unit cell. This also suggests that to achieve best linewidths under cryogenic conditions, the measurements of IP and OOP FMR need to be combined in order to identify the crystalline directions featuring the best linewidths, and to corroborate possible interpretations aiming to pinpoint the most detrimental loss mechanisms in a given film/substrate system. 

\section{Conclusion}
\label{sec:discuss}

We have investigated the dynamical magnetic properties of epitaxial thin films of YIG grown by LPE on a commercially available diamagnetic substrate, YSGG, and compared them to YIG films grown on GGG in similar conditions. Due to lattice mismatch, growing YIG on YSGG leads to biaxial tensile strain, which is well accommodated in films thinner than about \SI{100}{\nano\meter}. For thicker films, the strain starts to relax, which is accompanied by an abrupt degradation of the film quality. The use of YIG/YSGG eliminates the magnetic inhomogeneity and losses associated with substrate paramagnetism in GGG, which enables low FMR linewidths at cryogenic temperatures in any field orientation. The linewidths obtained for the present unpatterned films, reaching down to \SI{0.7}{\milli\tesla} at \SI{3}{\kelvin} and above \SI{40}{\giga\hertz}, are very promising for highly coherent magnonic or hybrid systems at low temperatures, and bring exciting opportunities for future quantum magnonics. We have identified different contributions to the FMR linewidth of YIG/YSGG in IP and OOP FMR conditions, which are related to substrate quality and tensile strain from the epitaxial growth. To enable further progress toward highest-quality single-crystalline films of iron garnet for cryogenic applications, it appears now crucial to pursue a careful investigation of the dissipation mechanisms in strained YIG, as well as in substituted iron garnets. The search for strategies to mitigate epitaxial strain in the growth of iron garnets on diamagnetic substrates is also very relevant to this goal, as demonstrated in very recent works investigating a new family of diamagnetic substrates lattice-matched with YIG\cite{Guguschev2025arXiv,Serha2025arXivB}, and lattice-tunable compositions of substituted iron garnets\cite{Legrand2025a}.

\clearpage
\begin{acknowledgments}
We acknowledge S.~Grenier for help in acquiring and interpreting the X-ray structural characterization data. This work has benefited from a government grant managed by the Agence Nationale de la Recherche as part of the France 2030 program, with reference ANR-24-EXSP-0005 (``MAGNON-BRAQET''). M.L.\ and R.S.\ acknowledge support from the Deutsche Forschungsgemeinschaft (DFG, German Research Foundation) via the SFB 1432, Project No.~425217212. M.L.\ additionally acknowledges support via Project No.~490730630.
\end{acknowledgments}

\appendix

\section{X-ray characterization}
\label{app:XRRD}

\begin{figure*}[b]
\includegraphics[width=7.0in, trim= 0cm 0cm 0cm 0cm]{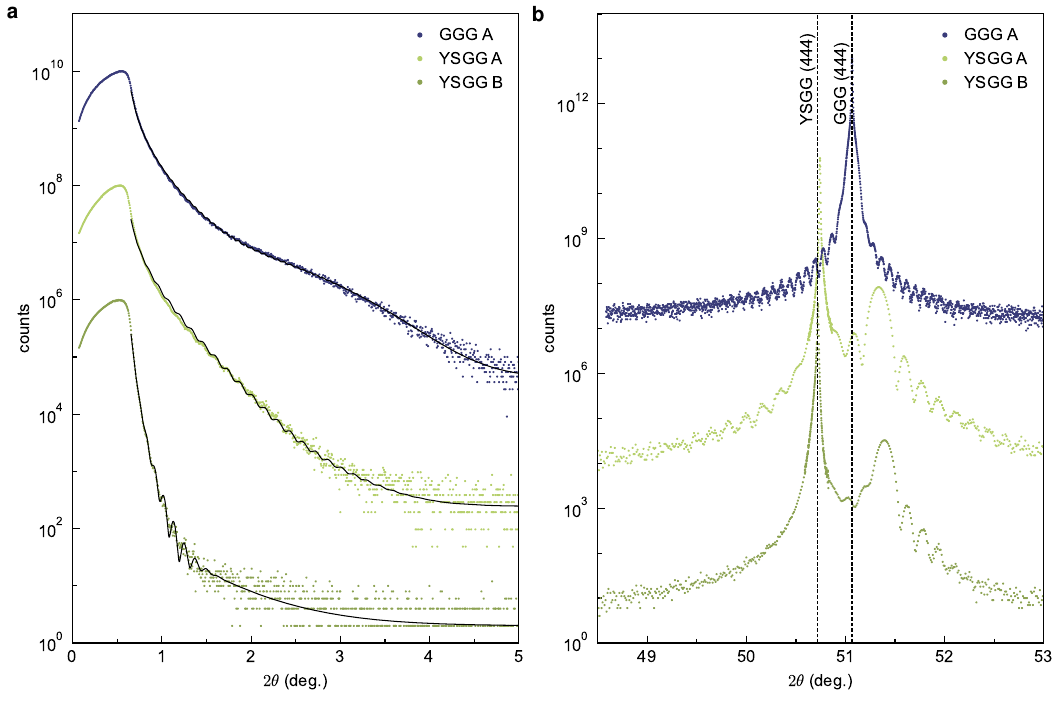}
\caption{(a) X-ray reflectivity curves and (b) high-resolution X-ray diffraction curves for samples GGG A, YSGG A and YSGG B. In panel a, the black lines are the reflectivity curves fit to the data (each sample offset by $10^2$). In panel b, the curves are $2\theta-\omega$ diffractograms near the (444) film peaks, with the dashed vertical lines locating the (444) diffraction peak of the substrates (each sample offset by $10^3$).}
\label{fig:XRRD}
\end{figure*}

X-ray reflectivity and diffraction curves have been acquired on a Rigaku SmartLab thin film diffractometer equipped with a Cu-tube, a parallel beam X-ray mirror, and a Ge(220) two-bounce Cu(K$\alpha$1) monochromator to suppress the Cu(K$\alpha$2,K$\beta$) radiation lines. The X-ray reflectivity curves are shown in Fig.~\ref{fig:XRRD}a, and are fitted to extract the YIG film thickness and roughness values listed in Table~\ref{tab:samples}. All the analysis of the present X-ray reflectivity curves has been performed in Python, relying on the package xrayutilities\cite{Kriegner2013}. The X-ray diffraction curves (symmetrical configuration, $2\theta-\omega$) are shown in Fig.~\ref{fig:XRRD}b. Nearby the (444) substrate peaks, the diffractograms display clear film peaks surrounded by marked Laue oscillations, corresponding to thickness fringes that confirm the high crystalline quality of the film growth. The positions of the substrate and film peaks provide the values of $a$ appearing in Table~\ref{tab:samples}.

\section{FMR measurement procedure}
\label{app:microwaves}

\begin{figure}[h]
\includegraphics[width=3.5in, trim= 0cm 0cm 0cm 0cm]{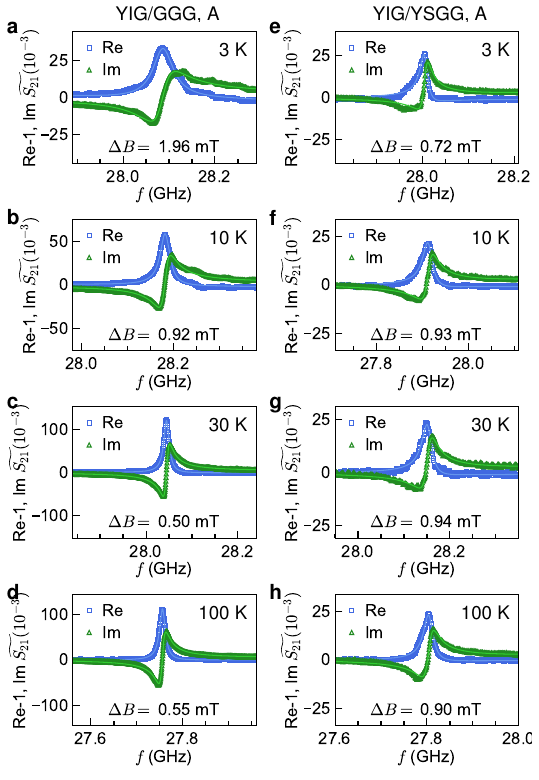}
\caption{Examples of microwave transmission parameter $\widetilde{S_{21}}(B_0)=S_{21}(B_0)/S_{21}(B_1)$, recorded for (a--d) YIG/GGG A and (e--h) YIG/YSGG A, at temperatures of (a,e) \SI{3}{K}, (b,f) \SI{10}{K}, (c,g) \SI{30}{K}, and (d,h) \SI{100}{K}. Each pair of curves corresponding to $\real{\widetilde{S_{21}}}(B_0)-1$ (blue squares) and $\imag{\widetilde{S_{21}}}(B_0)$ (green triangles) is fit with the microwave transmission model (blue and green lines).}
\label{fig:FMRsamplepeaks}
\end{figure}

The present broadband FMR measurements are based on the inductive coupling between the films and a microwave coplanar waveguide (CPW), mounted on a custom measurement rod fitted inside a variable temperature insert. The samples are placed on an impedance-matched constriction in the CPW line (width of the signal line is \SI{75}{\micro\meter}, each gap is \SI{140}{\micro\meter}). The microwave power is kept at \SI{-20}{dBm} or \SI{-10}{dBm} depending on linewidth, low enough to remain in a linear FMR excitation regime. The frequency-dependent microwave transmission data (scattering parameter $S_{21}$) around resonance at each external field step $B_0$ is normalized by the off-resonant data at a nearby field $B_1>B_0$\cite{Maier-Flaig2018,Legrand2025}, in order to minimize spurious signals associated to the variations of transmission with frequency and magnetic field inside the microwave components. Such $\widetilde{S_{21}}(B_0)=S_{21}(B_0)/S_{21}(B_1)$ traces can be fit by a model of the inductive microwave response of the sample at both $B_0$ and $B_1$. Examples of microwave transmission data $\widetilde{S_{21}}$ and fits for YIG/GGG A and YIG/YSGG A are provided in Fig.~\ref{fig:FMRsamplepeaks}. At each field value, $\real{\widetilde{S_{21}}}(B_0)-1$ and $\imag{\widetilde{S_{21}}}(B_0)$ are fit together to the microwave model to extract resonance frequency $f_{\rm{res}}$ and linewidth $\Delta{}B=\mu_0\Delta{}H$.

\section{Substrate stray field distribution}
\label{app:GGGfield}

\begin{figure}[h]
\includegraphics[width=3.5in, trim= 0cm 0cm 0cm 0cm]{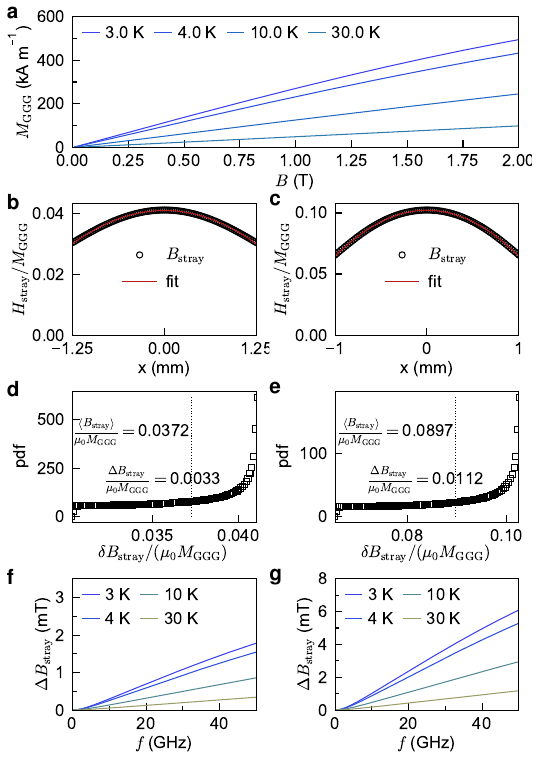}
\caption{Shape-dependent magnetic properties of GGG substrates: GGG A is \qtyproduct{2.5x4.0x0.5}{\milli\meter}, GGG B is \qtyproduct{2.0x2.0x0.5}{\milli\meter}. (a) In-plane magnetic field dependence of the magnetization in the GGG substrate, $M_{\rm{GGG}}$, at several temperatures (3, 4, 10 and \SI{30}{\kelvin}), for GGG A. Expected stray field profiles along the direction of the CPW probing line, aligned with $x$, obtained solely from geometrical considerations, for (b) GGG A and (c) GGG B. Resulting stray field normalized distribution (probability density function, pdf) for (d) GGG A and (e) GGG B. Contribution of the stray field distribution to the FMR linewidth, $\Delta{}B_{\rm{stray}}$, as a function of FMR resonance frequency $f$ at several temperatures (3, 4, 10 and \SI{30}{\kelvin}), for (f) GGG A and (g) GGG B.}
\label{fig:GGGprop}
\end{figure}

The influence of the paramagnetic substrate stray field on the FMR of thin films has been investigated in several works\cite{Danilov1989,Jermain2017,Mihalceanu2018,Roos2022,Knauer2023,Serha2024,Serha2025}. Here, we focus on the dependence on the substrate shape of the field inhomogeneity induced by the stray fields from the substrate. While the initial linear susceptibility $\chi_{\rm{p}}$ is sufficient to describe the paramagnetism of GGG at low fields and high temperatures, it otherwise does not keep its initial value $\chi_{\rm{p}} = \mu_0NJ(J+1)(g\mu_{\rm{B}})^2/(3k_{\rm{B}}T)$ with $N$ the density of \ce{Gd^{3+}} cations per unit volume, $J=7/2$ and $g=2$ for \ce{Gd^{3+}}, $\mu_{\rm{B}}$ the Bohr magneton, and $k_{\rm{B}}$ the Boltzmann constant. The paramagnetism of bulk GGG beyond its linear range can be approached by the Brillouin function $\mathcal{B}_{7/2}(x)=(8/7)\coth{(8/7)x}-(1/7)\coth{(1/7)x}$, considering $M_{\rm{GGG}}=Ng\mu_{\rm{B}}J\mathcal{B}_{J}(x)$ with $x=Jg\mu_{\rm{B}}/(k_{\rm{B}}T)$. The paramagnetism of the GGG substrates at low temperatures can be further calculated by a mean-field coefficients model\cite{Serha2024}, more accurate than a simple inverse temperature dependence for dense spin ensembles. To determine the substrate volumic moment at given temperature and field, as shown for several temperatures in Fig.~\ref{fig:GGGprop}a, it is required to determine the demagnetizing factors of the substrate\cite{Aharoni1998}, which reflect its lateral dimensions $a$, $b$, and $c$ across the $x,y,z$ directions. We can then solve\cite{Serha2024}
\begin{equation}
M_{\rm{GGG}}=M_{\rm{sat}}\cdot{}\mathcal{B}_{7/2}\left(\frac{7g\mu_B[B+(\lambda-\mathcal{N})\mu_0M_{\rm{GGG}}]}{2k_{\rm{B}}T}\right),
\end{equation}
with $M_{\rm{sat}}=$~\SI{813}{\kilo\ampere\per\meter} for GGG, $\mathcal{N}$ the relevant diagonal element of the demagnetizing tensor along the direction of the applied field, and $\lambda\approx-1.3$.

\begin{figure}[b]
\includegraphics[width=3.5in, trim= 0cm 0cm 0cm 0cm]{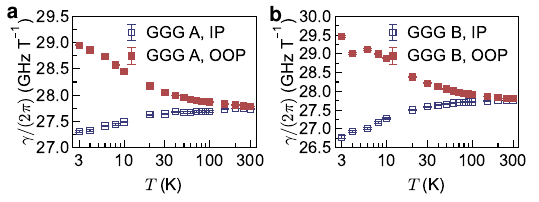}
\caption{Comparison of magnetization dynamics parameters obtained for IP FMR against OOP FMR, on films of YIG/GGG A and B, against temperature $T$. For (a) YIG/GGG A, and (b) YIG/GGG B, gyromagnetic ratio $\gamma/(2\pi)$. The difference originates in the shape-dependent magnetic properties of the stray field: GGG A is \qtyproduct{2.5x4.0x0.5}{\milli\meter}, GGG B is \qtyproduct{2.0x2.0x0.5}{\milli\meter}, creating different $\avg{B_{\rm{stray}}}$.}
\label{fig:gammaGGGOOP}
\end{figure}

For a given value of the substrate magnetization, a modeling\cite{Ortner2020} of the spatial profiles of the stray field (Fig.~\ref{fig:GGGprop}b,c) generated by the substrate also provides its distribution $\delta{}B_{\rm{stray}}$ (Fig.~\ref{fig:GGGprop}d,e). We checked that the error occurring from neglecting the non-uniform magnetization inside the substrate was negligible for the present purpose. The average of the computed field distribution $\avg{B_{\rm{stray}}}$ determines the predicted shift of the resonance field, and thus the effective $\gamma$ resulting from the sum of the external field and opposing (additional) stray field for an IP (OOP) external field\cite{Serha2024}. Those are shown in Fig.~\ref{fig:gammaGGGOOP}. As was found before, the deviation of $\gamma/(2\pi)$ from its room-temperature value is more pronounced for OOP than for IP, because the stray field inhomogeneity is higher when the GGG magnetization is aligned with a short direction of the substrate. Likewise, the shift of $\gamma/(2\pi)$ is larger for YIG/GGG B than for YIG/GGG A for geometrical reasons, as sample B has a smaller surface, and therefore causes significantly more inhomogeneous stray fields. 

We can then determine the contribution of this stray field distribution to the FMR linewidth\cite{Serha2025}, here by numerically computing its standard deviation $\Delta{}B_{\rm{stray}}$. Note that the field profiles are very well described by a quadratic spatial dependence, see fits in Fig.~\ref{fig:GGGprop}b,c. For a field deviation up to $-B_{\rm{diff}}$ on the samples edges compared to the stray field in the center of the sample, and approximating it by a quadratic dependence, the standard deviation is given by $\Delta{}B_{\rm{stray}}= 2B_{\rm{diff}}/(3\sqrt{5})$\cite{Legrand2025}. Using together (i) the inverse function of the Kittel law $B=([(\mu_0H_{\rm{eff}})^2+4(2\pi{}f/\gamma)^2]^{1/2}-\mu_0H_{\rm{eff}})/2$ to find the resonance field in a self-consistent way that includes its shift by $\avg{B_{\rm{stray}}}\propto{}M_{\rm{GGG}}(B,T)$; (ii) the deduced $M_{\rm{GGG}}(B,T)$; and (iii) the numerical determination of $\Delta{}B_{\rm{stray}}/(\mu_0M_{\rm{GGG}})$, finally yields the contribution of the stray field inhomogeneity to the FMR linewidth, $\Delta{}B_{\rm{stray}}$, as a function of frequency (Fig.~\ref{fig:GGGprop}f,g). The close proximity of the curves for $T=$~\SI{3}{\kelvin} and \SI{4}{\kelvin} highlights the onset of a saturated GGG magnetization at low temperatures and high fields. 

\bibliography{LPE_YIG_lowT}

\end{document}